# Ginzburg–Landau simulations of three-terminal operation of a superconducting nanowire cryotron

Naoki Yasukawa[1], Taichiro Nishio[1] and Yasunori Mawatari[2,*]

[1] Department of Physics, Tokyo University of Science, 1-3 Kagurazaka, Shinjuku, Tokyo 162-8601, Japan
[2] National Institute of Advanced Industrial Science and Technology (AIST), Tsukuba, Ibaraki 305-8568, Japan

Email: y.mawatari@aist.go.jp

**Abstract**

Superconducting nanowire cryotrons (nTrons) are expected to be used as interfaces for super-high-performance hybrid devices in which superconductor and semiconductor circuits are combined. However, nTrons are still under development, and diverse analyses of these devices are needed. Accordingly, we have developed a numerical technique to simulate the three-terminal operation of an nTron by using the finite element method to solve the time-dependent Ginzburg–Landau (TDGL) equation and the heat-diffusion equation. Simulations using this technique offer understanding of the dynamics of the order parameter, the thermal behavior, and the characteristics of three-terminal operation, and the TDGL model reproduces qualitatively the results of nTron experiments. In addition, we investigated how some geometric and physical parameters (the design elements) affect the operation characteristics. The TDGL model has far fewer free parameters compared with the lumped-element electrothermal model commonly used for simulating superconducting devices. Furthermore, the TDGL model provides time-dependent visual information about the superconducting state and the normal state, thereby offering insights into the relationship between nTron geometry and three-terminal operation. These simulation results offer a route to nTron optimization and the development of nTron applications.

## 1. Introduction

Superconducting single-flux-quantum (SFQ) circuits have remarkable features, such as a clock frequency of several hundred gigahertz and low power consumption [1–3]. However, the technology for the large-scale application of SFQ circuits is not yet established [4–6] because of their low integration density and low-voltage output signals, and it is difficult to create computing systems or memories with only SFQ and other superconducting technologies.

To facilitate high integration density, hybrid devices have been devised that combine SFQ circuits and complementary metal–oxide–semiconductor (CMOS) circuits [7,8]. These hybrid devices have the advantages of both SFQ circuits and CMOS circuits, such as high integration density and compatibility with conventional technologies and devices. However, the problem is that SFQ circuits and semiconductor circuits are not compatible with each other [9–11]. Since the output from an SFQ circuit is too small to drive a semiconductor circuit, an additional device is required as an interface to allow the two types of circuit to communicate with each other.

To compensate for the above disadvantage, three-terminal devices fabricated from a superconducting thin wire have been developed as amplifiers/discriminators [12, 13]. Subsequently, a new superconducting nanowire cryotron with a distinctive shape has been developed [14–16]. Known as a nanocryotron (nTron), its transition from the superconducting state to the normal state is used to amplify signals, and unusually for a superconducting device, the nTron has high output impedance. Its use as logic devices is also being considered [17]. However, to date only a few research groups [14–16] have used the nTron to develop hybrid SFQ–CMOS devices, and so the details of its operation mechanism are unclear, and there is still room for improving the nTron itself. Therefore, we numerically simulated the three-terminal operation of the nTron, investigating specifically how these important design elements of the nTron influence the operation characteristics. Simulation analyses offer both a basic understanding of the nTron operation mechanism and a path to modelling large-scale integrated circuits for developing hybrid devices.

The lumped-element electrothermal model and/or the two-temperature model has been commonly used to simulate superconducting-nanowire devices [12,14,15]. However, these models overlook the dynamics of vortices, and they do not account for the relaxation time of the order parameter. In contrast, our model allows for accurate simulation of the superconducting-normal transition phenomena with fewer parameters, while accounting for the characteristics mentioned above.

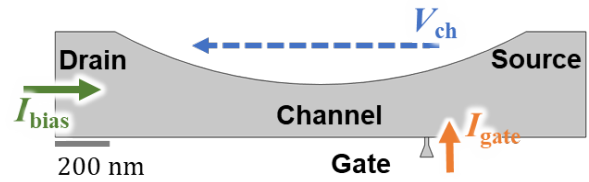

**Figure 1.** Schematic of nanocryotron (nTron) as modeled in simulation software.

This paper is organized as follows. In section 2, we present the simulation model and describe in detail what we use it to simulate. In section 3, we report the results of numerical calculations, and in section 4 we discuss our results in comparison to experimental results [14]. Finally, we present our conclusions in section 5.

## 2. Model

### 2.1. Modeled nTron and three-terminal operation

We designed the model geometry as shown in figure 1. The nTron has a source, a drain, and a gate terminal, and where the gate intersects the channel between the source and the drain is called the choke. Each size and positional relationship was decided based on previous work [14]. We simulated the following operation. Initially, the whole channel is in the superconducting state, and a finite gate current ($I_{\text{gate}}$) is applied to the gate. Then, a channel bias current ($I_{\text{bias}}$) is applied to the channel and is swept from 0 µA to 90 µA. Finally, after transition to the normal state occurs in the channel, the channel voltage ($V_{\text{ch}}$) is generated. The sweep rate of the bias current is set to $dI_{\text{bias}}/dt = 10^4$ A/s. If the sweep rate is too large ($dI_{\text{bias}}/dt \gtrsim 10^5$ A/s) compared to the rate at which the normal-state region in the channel expands, the numerical $V_{\text{ch}}$ is much smaller than the experimental value. If the sweep rate is too small ($dI_{\text{bias}}/dt \lesssim 10^3$ A/s), on the other hand, the computational cost becomes too large. We thus choose $dI_{\text{bias}}/dt = 10^4$ A/s as the optimal value to simulate the experimental data.

*2.2. Governing equations, material properties, and other conditions*

We numerically simulated the three-terminal operation of the nTron, mainly studying its $I_{\text{bias}}$–$V_{\text{ch}}$ characteristics. We numerically solved the time-dependent Ginzburg–Landau (TDGL) equation and the heat-diffusion equation [18,19] using the finite element method. The dynamics of the nTron's order parameter $\psi$ are described by the TDGL equation [20],

$$\tau_{\text{GL}}\left(\frac{\partial}{\partial t} + i\frac{2\pi}{\phi_0}\varphi\right)\psi - \xi^2 \nabla^2 \psi + \left(|\psi|^2 + \frac{T}{T_c} - 1\right)\psi = 0, \quad (1)$$

where $\tau_{\text{GL}} = h/(16 k_B T_c)$ $(= 0.25\text{ ps})$ is the relaxation time, $\xi$ is the coherence length, $\phi_0$ is the flux quantum, $\varphi$ is the electrical scalar potential, $T$ is the electron temperature, and $T_c$ is the critical temperature. No magnetic field is applied, and we neglect the vector potential $A$ in the TDGL equation. The self-field due to the current in the nTron is also neglected because the nTron is composed of a nanostrip of width $w_{\max} < \lambda^2/d$ (where $\lambda$ is the penetration depth, $w_{\max}$ is the maximum channel width of the nTron, and $d$ is the thickness of the film). We also need an equation for current conservation, that is,

$$\frac{\nabla^2 \varphi}{\rho_n} = \nabla \cdot \boldsymbol{j}_s, \quad (2)$$

where $\rho_n$ is the normal state resistivity and

$$\boldsymbol{j}_s = \frac{\phi_0}{4 i \pi \mu_0 \lambda^2}(\psi^* \nabla \psi - \psi \nabla \psi^*),$$

is the superconducting current density, with $\mu_0$ being the vacuum permeability.

The nTron's thermal behavior is found from the heat-diffusion equation, that is,

$$C\frac{\partial T}{\partial t} - K\nabla^2 T + \frac{C}{\tau_{\text{el-ph}}}(T - T_{\text{bath}}) = P_{\text{Joule}} + P_{\text{opr}}, \quad (3)$$

where $C$ is the heat capacity, $K$ is the thermal conductivity, $\tau_{\text{el-ph}}$ is the electron–phonon relaxation time, $T_{\text{bath}}$ is the bath temperature,

$$P_{\text{Joule}} = |\boldsymbol{E}|^2 / \rho_n,$$

is the Joule dissipation power, $\boldsymbol{E} = -\nabla \varphi$ is the electric field, and

$$P_{\text{opr}} = \frac{1}{\mu_0 \tau_{\text{GL}}}\left(\frac{\phi_0}{2\pi\xi\lambda}\right)^2 \left|\tau_{\text{GL}}\left(\frac{\partial}{\partial t} + i\frac{2\pi}{\phi_0}\varphi\right)\psi\right|^2,$$

is the dissipation power due to relaxation of the order parameter [20].

In addition, our simulations have accounted for the effects of nTron inductance and load resistance. We coupled the following circuit equation,

$$L_k \frac{dI_{\text{ch}}}{dt} + V_{\text{ch}} = R_L(I_{\text{bias}} - I_{\text{ch}}), \quad (4)$$

where $L_k$ is the kinetic inductance of the channel, $I_{\text{ch}}$ is the current flowing through the channel, and $R_L$ is the load resistance. $V_{\text{ch}}$ is calculated from the electrical scalar potential $\varphi$.

We used the time-dependent solver of COMSOL Multiphysics® [21] to solve these partial differential equations subject to the following boundary conditions. First, the superconducting current does not flow through the perimeter edges of the nTron, giving $\mathbf{n} \cdot \nabla \psi = 0$, where $\mathbf{n}$ is the normal vector at the edges.

**Table 1.** Simulation conditions and physical parameters of thin film of niobium nitride [14,22,23].

| $\tau_{\text{GL}}$ [ps] | $\xi$ [nm] | $T_c$ [K] | $C$ [J/m$^3$/K] | $d$ [nm] | $\tau_{\text{el-ph}}$ [ps] |
|---|---|---|---|---|---|
| 0.25 | 7.5 | 12 | 2400 | 10 | 17 |
| $\lambda$ [nm] | $w_{\max}$ [nm] | $T_{\text{bath}}$ [K] | $K$ [W/m/K] | $L_k$ [nH] | $R_L$ [kΩ] |
| 400 | 400 | 4.2 | 0.108 | 1 | 10 |

$\tau_{\text{GL}}$—relaxation time, $\xi$—coherence length, $T_c$—critical temperature, $C$—heat capacity, $K$—thermal conductivity, $\tau_{\text{el-ph}}$—electron–phonon relaxation time, $\lambda$—penetration depth, $w_{\max}$—maximum channel width of nTron, $d$—thickness of film, $T_{\text{bath}}$—bath temperature, $L_k$—the kinetic inductance of the channel, $R_L$—the load resistance.

**Table 2.** Design elements and their standard values (see figure 2).

| $P_g$[nm] | $W_g$[nm] | $W_c$[nm] | $\rho_n$[μΩm] |
|---|---|---|---|
| 400 | 15 | 190 | 2.85 |

$P_g$—distance between gate and center of channel, $W_g$—size of choke (intersection of channel and gate), $W_c$—constriction width of channel, $\rho_n$—resistivity in normal state.

Second, the normal current does not flow through the perimeter edges of the nTron except at the three terminals, giving $\mathbf{n} \cdot \nabla \varphi = 0$. Finally, no heat flows through the perimeter edges of the nTron, giving $\mathbf{n} \cdot \nabla T = 0$. Table 1 gives the simulation conditions and the physical parameters of the thin film of niobium nitride (NbN) used in our simulation.

### 2.3. Design elements

As previously mentioned, we focused on understanding the effects of the nTron's geometric and physical parameters. Table 2 shows the design elements, which comprise three geometric dimensions and one physical property. Their standard values were set using values extracted from scanning electron micrographs of the nTron [14]. In order to investigate how they affect the three-terminal operation of the nTron, we simulated its behavior under different values of the design elements as follows.

*Position of gate* ($P_g$). The first design element is the position of the gate, which is defined as the distance between the gate and the center of the channel. This element plays an important role in the nTron's operating principle because transition to the normal state occurs easily at the intersection of the channel and the gate. The standard value was $P_g$ = 400 nm, but we also simulated the nTron behavior with $P_g$ = 0 nm.

*Size of choke* ($W_g$). The next design element is size of the choke. The choke is the smallest nTron component, and its size influences the superconducting-to-normal transition of the gate. The standard value was $W_g$ = 15 nm, but we also simulated the nTron behavior with $W_g$ = 10 nm.

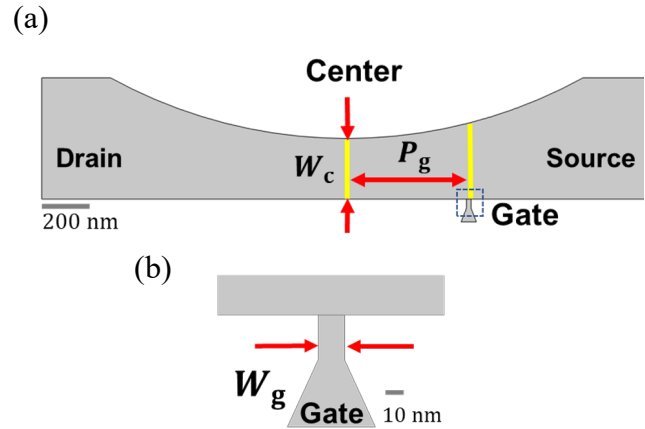

**Figure 2.** Schematic of design elements: (a) $P_g$ is the distance between the gate and the center of the channel, and $W_c$ is the constriction width of the channel; (b) close-up of choke, which is the dashed box in (a), with $W_g$ being the size of the choke, which is the intersection of the channel and the gate.

*Constriction width* ($W_c$). The third design element is the constriction width of the channel. The critical current density is smallest in the channel, and without the gate current, the normal transition begins to occur at the constriction. The standard value was $W_c$ = 190 nm, but we also simulated the nTron behavior with $W_c$ = 150 nm.

*Normal state resistivity* ($\rho_n$). The final design element is the normal state resistivity. The value of $\rho_n$ has a direct influence on the critical current of the constriction and the choke. The standard value was $\rho_n$ = 2.85 μΩm, but we also simulated the nTron behavior with $\rho_n$ = 10 μΩm.

### 3. Results

In this section, we present simulations results for five scenarios, along with their corresponding explanations.

*3.1. nTron designed with values in Tables 1 and 2*

Figure 3 shows the simulation results for the nTron designed with the parameter values given in tables 1 and 2. The graph shows the $I_{bias}$–$V_{ch}$ characteristics for $I_{gate}$ values from 0 μA to 10 μA, with the horizontal axis representing the channel bias current $I_{bias}$ and the vertical axis representing the channel voltage $V_{ch}$. For small $I_{gate}$ (0 μA to 5 μA), the $I_{bias}$–$V_{ch}$ curves overlap, and the $I_{bias}$ value when $V_{ch}$ begins to rise is $I_{c,bias} = 83.0$ μA, where $I_{c,bias}$ is defined as the $I_{bias}$ value when $V_{ch} = 1$ mV. The time it took for the channel voltage to reach $V_{ch} = 100$ mV is typically ∼100 ps. For $I_{gate} \geq 6$ μA, as $I_{gate}$ increases, $I_{c,bias}$ decreases systematically; specifically, for $I_{gate} = 6$ μA, 7 μA, 8 μA, 9 μA, and 10 μA, we have $I_{c,bias}$= 76.3 μA, 71.3 μA, 67.3 μA, 64.0 μA, and 61.2 μA, respectively. What all the $I_{bias}$–$V_{ch}$ curves have in common is that their slope becomes gradual after $V_{ch}$ rises up almost vertically. When the transition to the normal state starts to occur in the channel, the channel voltage rises sharply, and there is the point at which the channel voltage begins to grow less slowly. After the rapid voltage rise, the $I_{bias}$–$V_{ch}$ curve exhibits essentially Ohmic behavior. This differing steepness of the channel voltage is explained in more detail in section 3.2.

Figure 4 shows visualization snapshots of the transition from the superconducting state to the normal state. They show the normalized super-electron density $|\psi|^2$ on a color scale, with blue corresponding to the superconducting state and red corresponding to the normal state. Figure 4(a) – (e) shows the distribution of $|\psi|^2$ for $I_{gate} = 5$ μA and $I_{bias} =$ (a) 80.0 μA, (b) 82.9 μA, (c) 83.0 μA, (d) 83.2 μA, and (e) 83.9 μA. When $I_{bias} =$ 80.0 μA and $V_{ch} \approx 0$ mV, the normal transition does not occur in the channel [figure 4(a)], but when $I_{bias} = 82.9$ μA and $V_{ch} = 0.1$ mV, the transition occurs slightly at the constriction of the channel [figure 4(b)]. As $I_{bias}$ increases further, the area in the normal state expands from the center to both ends of the channel. Similar simulation results were obtained for $I_{gate} \leq 4$ µA. This is one of the reasons why the $I_{bias}$–$V_{ch}$ characteristics for $I_{gate}$ = 0 µA to 5 µA show the same results. Figure 4(f)–(j) shows the distribution of $|\psi|^2$ for $I_{gate} = 6$ µA. When $I_{bias} = 60.0$ µA and $V_{ch} \approx 0$ mV [figure 4(f)], the normal transition occurs only at the vicinity of the gate. As $I_{bias}$ increases [(g) 76.3 µA, (h) 76.8 µA, (i) 77.3 µA, and (j) 79.0 µA], the area in the normal state extends across the channel and finally expands to the whole channel. Similar simulation results were obtained for $I_{gate} \geq 7$ µA.

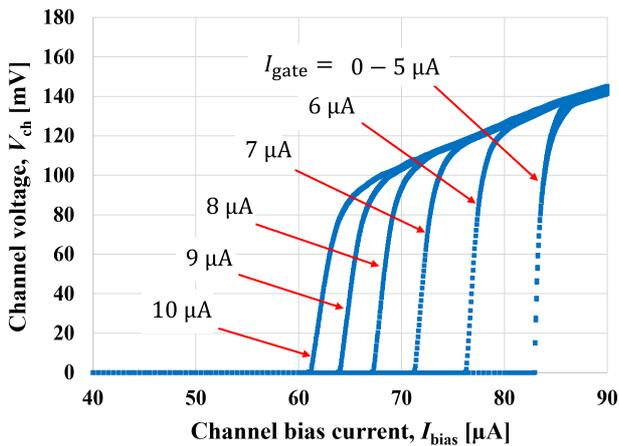

**Figure 3.** Characteristics of channel bias current ($I_{bias}$) versus channel voltage ($V_{ch}$) for $I_{gate} = 0$ μA to 10 μA. The design elements have the values given in tables 1 and 2.

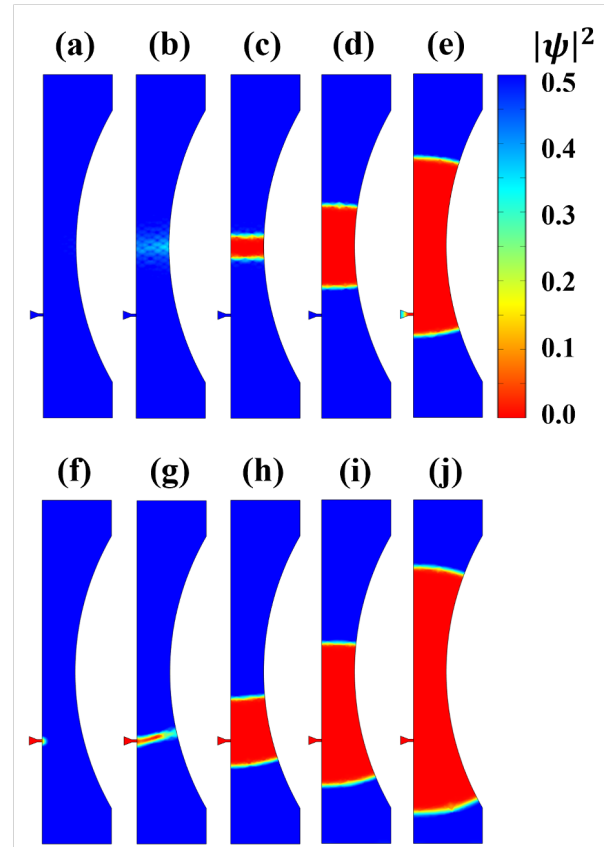

**Figure 4.** Distribution of normalized super-electron density $|\psi|^2$ for (a)–(e) $I_{gate} = 5$ μA and for (f)–(j) $I_{gate} = 6$ μA.

### 3.2. nTron designed with $P_g = 0\ nm$

Figure 5 shows the simulation results obtained under the same conditions as those in section 3.1. except for the gate position, which is now $P_g = 0$ nm. Unlike in the results shown in figure 3, each $I_{bias}$–$V_{ch}$ curve for $I_{gate} \leq 5$ µA can now be distinguished. For $I_{gate} = 0$ µA to 4 µA, $I_{c,bias}$ is around 80 µA. For $I_{gate} = 5$ µA to 10 µA, all the $I_{c,bias}$ values are smaller than those for $P_g = 400$ nm shown in section 3.1. This $P_g$ dependence of $I_{c,bias}$ is reasonable because the total current of $I_{bias} + I_{gate}$ flows at the constriction of the channel when $P_g = 0$ nm.

As mentioned in section 3.1, for small $I_{gate}$ ($I_{gate} = 0$ µA to 4 µA) $V_{ch}$ increases almost vertically with $I_{bias}$, whereas for large $I_{gate}$ ($I_{gate} = 5$ µA to 10 µA), $V_{ch}$ increases rather gradually. Because this behavior is seen regardless of the gate position, the main reason has no relation to the channel geometry. When the channel voltage increases almost vertically, sweeping the channel bias current mainly causes the normal transition. Once the normal transition occurs at the center of the channel, the current flowing through both sides of the center easily reaches the critical current, and these areas immediately switch to the normal state. In contrast, when the channel voltage rises somewhat gradually, the major factor causing the transition to the normal state is the gate current at the initial stage. In other words, the area around the gate enters the normal state, and this temporarily decreases the channel width where the supercurrent flows. Indeed, the transition to the normal state occurs in the vicinity of the gate, but the current flowing through both sides of the gate is insufficient to cause the subsequent transition. This is because both sides of the gate have the original channel width.

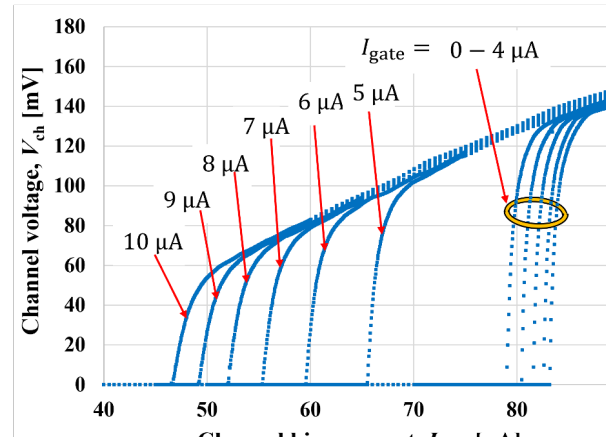

**Figure 5.** $I_{bias}$–$V_{ch}$ characteristics for $I_{gate}$ = 0–10 µA. The simulation conditions are the same as those for figure 3 except for the gate position, which is now $P_g = 0$.

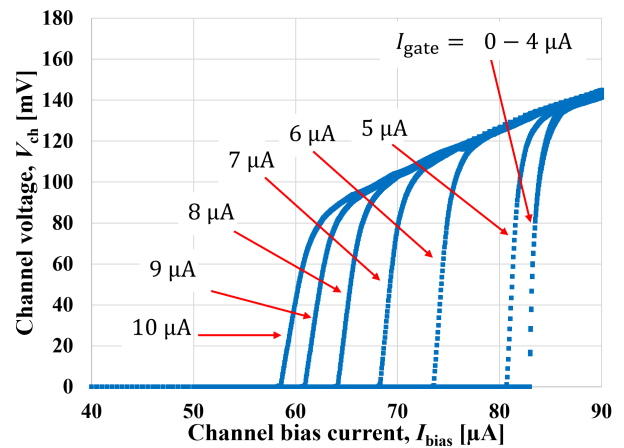

**Figure 6.** $I_{bias}$–$V_{ch}$ characteristics for $I_{gate}$ = 0 µA to 10 µA. The simulation conditions are the same as those in figure 3 except for the choke size $W_g$, which is now 10 nm.

*3.3. nTron designed with $W_g = 10\ nm$*

Figure 6 shows the simulation results for the smaller choke width $W_g = 10$ nm. Unlike in the results for $W_g = 15$ nm shown in figure 3, the $I_{bias}$–$V_{ch}$ curve for $I_{gate} = 5$ μA can now be distinguished from those for $I_{gate} = 0$ - 4 μA. For $I_{gate} = 6$ μA to 10 μA, the $I_{c,bias}$ values for $W_g = 10$ nm shown in figure 6 are smaller than those for $W_g = 15$ nm shown in figure 3. Reducing the choke size decreases the critical current of the gate, which may then heat up more easily, thereby reducing $I_{c,bias}$.

*3.4. nTron designed with $W_c = 150\ nm$*

Figure 7 shows the simulation results for the smaller channel constriction width $W_c = 150$ nm. Although the overall behavior of the $I_{bias}$–$V_{ch}$ curves is qualitatively similar to that shown in figure 3 for $W_c = 190$ nm, each value of $I_{c,bias}$ is smaller. For example, for $I_{gate} = 10$ μA, we have $I_{c,bias} = 49.0$ μA and 61.2 μA for the nTron with $W_c = 150$ nm and 190 nm, respectively. The ratio of the two different constriction widths, that is, (150 nm)/ (190 nm) = 0.79, is close to that of the two $I_{c,bias}$ values, that is, (49.0 μA)/ (61.2 μA) = 0.80. $I_{c,bias}$ is directly proportional to $W_c$. Moreover, compared with the results in section 3.1., the reachable $V_{ch}$ value with the same $I_{bias}$ is higher. It is reasonable for the nTron to have higher resistance because the resistance is determined by $\rho_n L/W$ (where $L$ is the channel length and $W$ is the channel width in the normal state). This might be an important element in view of application to the amplifiable interface.

*3.5. nTron designed with $\rho_n = 10\ \mu\Omega m$*

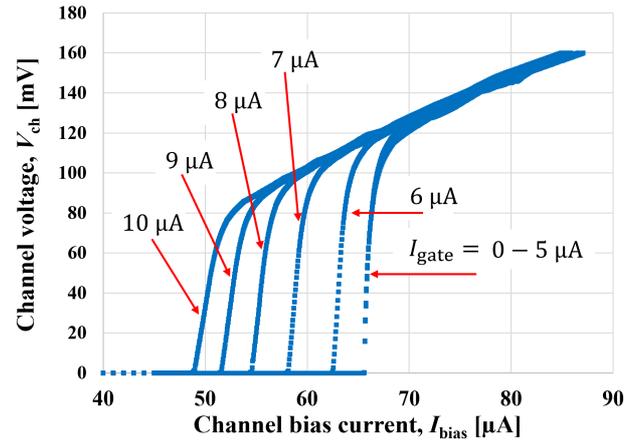

**Figure 7.** $I_{bias}$–$V_{ch}$ characteristics for $I_{gate}= 0$ μA to 10 μA. The simulation conditions are the same as those in figure 3 except for the constriction width $W_c$, which is now 150 nm.

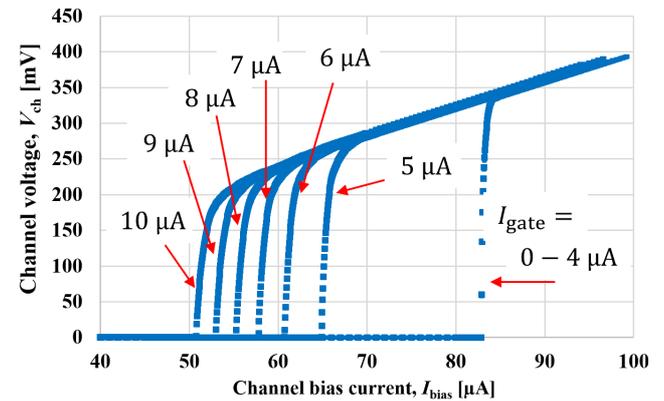

**Figure 8.** $I_{bias}$ –$V_{ch}$ characteristics for $I_{gate} = 0$ μA to 10 μA. The simulation conditions are the same as those in figure 3 except for the normal resistivity $\rho_n$, which is now 10 μΩm.

Figure 8 shows the simulation results for the larger normal-state resistivity of $\rho_n = 10$ μΩm. Here, the value of $I_{c,bias} = 82.9$ μA for $I_{gate} = 0$ to 4 μA is close to that from figure 3. The result for $I_{gate} = 5$ μA is different from the results for $I_{gate}$ 0 μA to 4 μA. The $I_{c,bias}$ values for $I_{gate} = 6$ μA to 10 μA are smaller than those of the results in figure 3. The overall $V_{ch}$ values are larger than those in figure 3.

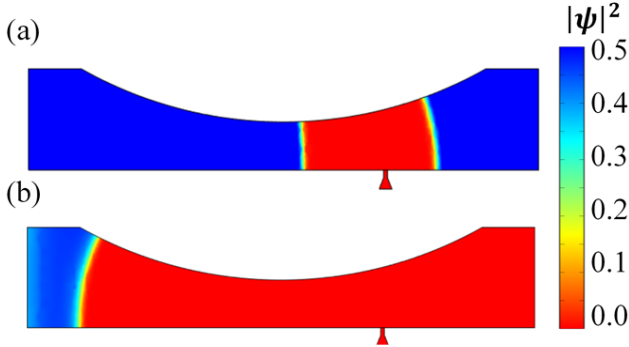

**Figure 9.** Distribution of normalized super-electron density $|\psi|^2$ for $I_\text{gate} = 9$ µA and $I_\text{bias} = 65$ µA with (a) $\rho_\text{n} = 2.85$ µΩm and (b) $\rho_\text{n} = 10$ µΩm.

Figure 9 shows the distribution of the super-electron density for $I_\text{gate} = 9$ µA and $I_\text{bias} = 65.0$ µA with (a) $\rho_\text{n} = 2.85$ µΩm and $V_\text{ch} = 40.6$ mV and (b) $\rho_\text{n} = 10$ µΩm and $V_\text{ch} = 263$ mV. This shows that the nTron with higher $\rho_\text{n}$ generates higher $V_\text{ch}$.

## 4. Discussion

We compared our simulation results shown in figure 3 with the experimental data in Ref. [14]. Although our simulation results reproduce the experimental data qualitatively, the $I_\text{bias}$–$V_\text{ch}$ curves for small $I_\text{gate}$ show different behavior. In the experimental data of figure 1C in Ref. [14], every $I_\text{gate}$ value can control the $I_\text{bias}$–$V_\text{ch}$ characteristics. One of the reasons for this difference is our supposition about physical properties. In our simulation, we supposed that the entire region of the nTron has uniform physical properties, but the real nTron might have inhomogeneous ones. Indeed, in the nTron fabrication process, partial degradation of the superconducting properties and inhomogeneity of the normal resistivity can occur unintentionally. For example, an nTron whose choke has high normal resistivity might be more sensitive to small $I_\text{gate}$.

Note that $\rho_\text{n}$ strongly affects the $I_\text{bias}$–$V_\text{ch}$ characteristics for only large $I_\text{gate}$. Because higher $\rho_\text{n}$ causes more heat generation, $I_\text{c,bias}$ becomes small, and for small $I_\text{gate}$, until the transition to the normal state occurs at the constriction of the channel, the area in the normal state does not exist. For that reason, these $I_\text{c,bias}$ values are unaffected by $\rho_\text{n}$. Using superconducting films with high $\rho_\text{n}$ might be advantageous regarding the operating speed: the nTron with $\rho_\text{n} = 2.85$ µΩm and $I_\text{gate} = 10$ µA took 720 ps to reach $V_\text{ch} = 100$ mV, whereas that with $\rho_\text{n} = 10$ µΩm and $I_\text{gate} = 10$ µA took only 70 ps to reach the same $V_\text{ch}$. Therefore, choosing appropriate materials facilitates faster operations without degrading the properties.

We employ the finite element method, enabling us to modify the geometry of the nTron. Hence, we can optimize the nTron by changing elements such as the choke and/or channel edges. Furthermore, the Ginzburg-Landau (GL) equation can easily handle models with spatially inhomogeneous superconducting properties, such as the critical temperature $T_\text{c}$ in Eq. (1). By utilizing the advantages of the simulation methods and the governing equations as mentioned above, we aim to optimize devices by addressing the inhomogeneities that appear in real devices.

While the TDGL model is capable of accurate simulation, it has the disadvantage of long computation time. For instance, the total computation time was ~3 days to obtain 11 curves of $I_\text{bias} - V_\text{ch}$ characteristics in figure 3. The simulation requires a relatively short computation time when the nTron is in the superconducting state. In contrast, a considerably longer computation time is needed for simulations when the normal region of the nTron extends.

## 5. Conclusion

We numerically simulated the three-terminal operation of an nTron by solving the TDGL

equation and the heat-diffusion equation, and from the results we draw the following conclusions.

The center gate position improves the controllability of $I_{c,\text{bias}}$ without sacrificing the output characteristics, and this might allow the nTron to process input signals of various intensities. Also, choosing suitable $I_{c,\text{bias}}$ is important for reducing power consumption. In addition, high normal resistivity improves the response speed and output voltage; while it might be difficult to design the normal resistivity arbitrarily, it is worth taking it into account.

Numerical simulations based on the TDGL and heat-diffusion equations can be used to optimize further the nTron's geometry and other parameters. In addition, nTrons are expected to be used for amplifying the pulse signals from SFQ circuits, and simulation of the pulse current operation will be performed. Also, it is not realistic to simulate more complex circuits with nTrons by using the TDGL model because of the large amount of computation time required. Therefore, we are planning to use a SPICE simulator for circuit simulations, and the results of this study can be effectively used to build SPICE models of the nTron. These simulations and analyses offer understanding of nTron operations and help in developing ultra-high-performance hybrid SFQ–CMOS systems.

**Acknowledgments**

We thank N. Yoshikawa and Y. Yamanashi for helpful discussions about nTron operations. This work was supported by JSPS KAKENHI Grant Number JP20K05314.

**Appendix**

*A.1. A hysteretic $I_{\text{bias}} - V_{\text{ch}}$ curve*

We investigated the hysteretic $I_{\text{bias}} - V_{\text{ch}}$ characteristics when the bias current was swept up and then down. Specifically, we swept $I_{\text{bias}}$ at a sweep rate of $dI_{\text{bias}}/dt = 10^4$ A/s until $I_{\text{bias}} = 86$ μA, and then reduced the $I_{\text{bias}}$ at the same rate.

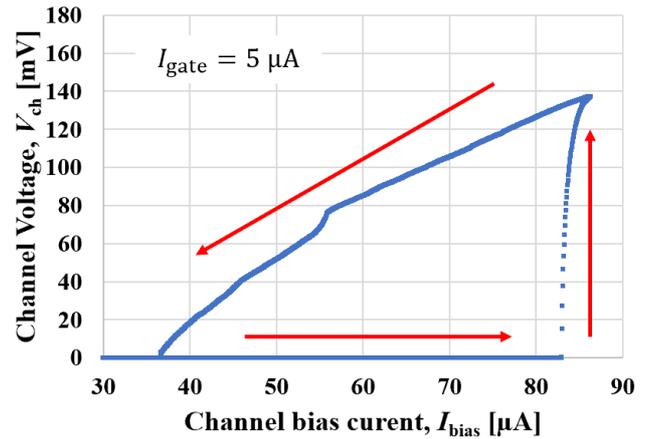

**Figure A1.** A hysteretic $I_{\text{bias}}$–$V_{\text{ch}}$ characteristic for $I_{\text{gate}}= 5$ μA. The $I_{\text{bias}}$ is swept until $I_{\text{bias}} = 86$ μA and swept down at the same rate.

Figure A1 shows the simulation result for the nTron designed with the parameter values given in tables 1 and 2. The gate current is 5 μA. $V_{\text{ch}}$ begins to develop at $I_{\text{bias}} = 83$ μA and rises sharply to around $V_{\text{ch}} = 140$ mV at $I_{\text{bias}} = 86$ μA. Thereafter, as $I_{\text{bias}}$ decreases, $V_{\text{ch}}$ decreases gradually. $V_{\text{ch}}$ becomes zero again when $I_{\text{bias}} = 36$ μA.

*A.2. Effects of a load resistance*

We investigated the effect on the $I_{\text{bias}} - V_{\text{ch}}$ characteristics when the load resistance $R_{\text{L}}$ is changed. Figure A2 shows the simulation results for the nTron designed with the parameter values given in tables 1 and 2 except for $R_{\text{L}}$. The load resistance is $R_{\text{L}} = 50$ Ω and 500 Ω, and the gate current is $I_{\text{gate}} = 5$ μA in both cases. When $R_{\text{L}} = 50$ Ω, $V_{\text{ch}}$ is generated at $I_{\text{bias}} = 83$ μA, it does not rise significantly (i.e., $V_{\text{ch}} < 3$ mV even when $I_{\text{bias}} = 90$ μA). This is because a large portion of $I_{\text{bias}}$ can easily flow into the load resistor when the normal transition in the channel occurs. On the other hand, when $R_{\text{L}} = 500$ Ω, the $I_{\text{bias}} - V_{\text{ch}}$ characteristic is qualitatively the same as in figure 3 for $R_{\text{L}} = 10$ kΩ. After $V_{\text{ch}}$ is generated at $I_{\text{bias}} = 83$ μA, it rises sharply and then continues to rise gently. Compared to the result in figure 3, the absolute value of $V_{\text{ch}}$ is small.

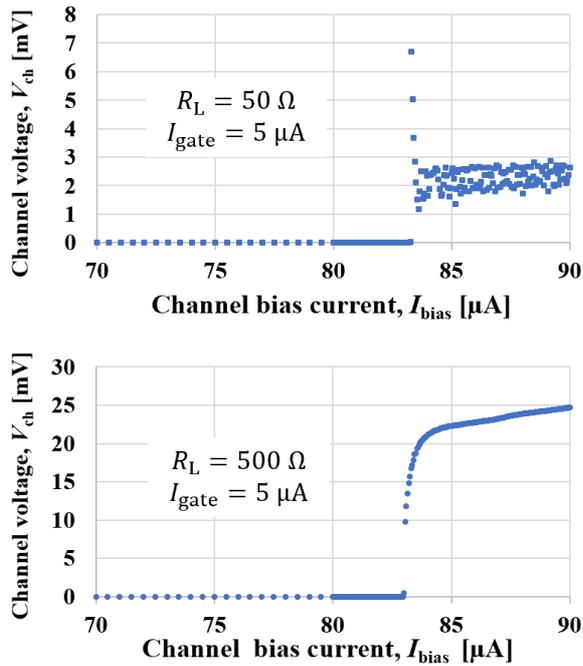

**Figure A2.** $I_{bias}$–$V_{ch}$ characteristics for $I_{gate}$ = 5 μA. The simulation conditions are the same as those in figure 3 except for $R_L$ = 50 Ω (upper panel) and $R_L$ = 500 Ω (lower panel).

**Referenes**